# Controlling the plasmonic properties of ultrathin TiN films at the atomic level

*Deesha Shah[1], Alessandra Catellani[2], Harsha Reddy[1], Nathaniel Kinsey[3], Vladimir Shalaev[1], Alexandra Boltasseva[1], Arrigo Calzolari[2*]*

[1] School of Electrical and Computer Engineering and Birck Nanotechnology Center, Purdue University, West Lafayette, Indiana 47907, USA

[2] CNR-NANO Istituto Nanoscienze, Centro S3, I-41125 Modena, IT

[3] School of Electrical and Computer Engineering, Virginia Commonwealth University, Richmond, Virginia 23220, USA

* arrigo.calzolari@nano.cnr.it

**ABSTRACT**

By combining first principles theoretical calculations and experimental optical and structural characterization such as spectroscopic ellipsometry, X-ray spectroscopy, and electron microscopy, we study the dielectric permittivity and plasmonic properties of ultrathin TiN films at an atomistic level. Our results indicate a remarkably persistent metallic character of ultrathin TiN films and a progressive red shift of the plasmon energy as the thickness of the film is reduced. The microscopic origin of this trend is interpreted in terms of the characteristic two-band electronic structure of the system. Surface oxidation and substrate strain are also investigated to explain the deviation of the optical properties from the ideal case. This paves the way to the realization of ultrathin TiN films with tailorable and tunable plasmonic properties in the visible range for applications in ultrathin metasurfaces and flexible optoelectronic devices.

**KEYWORDS** plasmonics, optical properties, DFT, ellipsometry, titanium nitride

**1. INTRODUCTION**

Recent developments in nanofabrication techniques have facilitated a surge of nanoplasmonic devices whose properties can be engineered by careful structural control of their metallic building blocks[1]. The usage of plasmonic materials – mostly metals – that support the light-coupled subwavelength oscillations of free electron clouds has led to advances in various applications, such as sensing[2–4], photovoltaics[5–7], and optical circuitry[8,9], unachievable with conventional dielectric photonic materials. Along with the control of the lateral dimensions of plasmonic structures, it is also possible to tailor the properties by adjusting the thickness of plasmonic materials, all the way down to a few monolayers[10,11]. As the dimensions shrink down to nanometer range thicknesses, quantum phenomena emerge as a result of the strong electron confinement, making plasmonic materials ideal platforms to study light – matter interactions at the nanoscale[12,13]. For example, the strong confinement leads to an enhanced nonlinear optical response in ultrathin films in comparison to their bulk counterparts[14,15]. Most importantly, ultrathin plasmonic films are predicted to have a greatly increased sensitivity to the local dielectric environment, strain, and external perturbations[10]. Consequently, unlike conventional metals with properties that are challenging to tailor, atomically thin plasmonic materials exhibit optical responses that can be engineered by precise control of their thickness, composition, and the electronic and structural properties of the substrate and superstrate[10,16,17]. This unique tailorability, unachievable with bulk or relatively thick metallic layers, establishes ultrathin plasmonic films as an attractive material platform for the design of tunable and dynamically switchable metasurfaces. In order to realize these applications, a deeper insight into plasmonic material properties at the nanoscale is required.

To investigate the sought after atomically-thin plasmonic regime, there have been several efforts to grow ultrathin (<10nm) metallic films with various plasmonic materials, such as gold[18] and titanium nitride (TiN)[11,19]. For noble metals, which are the usual choice of material for plasmonic structures, the deposition of smooth, continuous ultrathin films is very challenging due to island formation and large defect concentrations[18,20]. In contrast, the epitaxial growth of TiN on lattice matched substrates, such as MgO, allows for the growth of thin films with thicknesses down to 2 nm while preserving their metallic properties[21,22]. To move towards the realization of practical devices utilizing ultrathin plasmonics, additional studies are needed to understand the optical behavior of ultrathin films and the different factors contributing to their thickness dependent properties. In this study, we present both a theoretical and an experimental study on the dielectric permittivity of ultrathin TiN films of varying thicknesses. We show that the effects of oxidation and strain on the optoelectronic and plasmonic properties of the thin films emerge as fundamental parameters to optimize the nanostructure response.

## 2. METHOD

<u>Theory</u>: First principles density functional (DFT) calculations are carried out with semilocal (PBE[23]) exchange-correlation functional, using plane wave basis set (with a cutoff energy of 200 Ry) and norm-converging pseudopotentials[24], as implemented in Quantum Espresso[25] codes. A uniform (28x28) k-point grid is used for summations over the 2D Brillouin zone.

Extended TiN films are simulated using periodically repeated supercells. Each unit cell has a (√2 x √2) 2D lateral periodicity and contains 1 to 10 layers of TiN(001), as shown in Figure 1a. The lattice parameter of TiN surface (4.24 Å) is obtained from the optimization of the corresponding bulk crystal[26]. When considering the effect of oxidation, O adatoms are symmetrically adsorbed on each surface to avoid a spurious dipolar field across the cell. Slab replicas are separated by at least 30 Å of vacuum. Each structure is fully relaxed until forces on all atoms become lower than 0.03 eV/Å.

The complex dielectric function $\hat{\varepsilon} = \varepsilon' + i\varepsilon''$ is evaluated using the code epsilon.x, also included in the Quantum ESPRESSO suite. This code implements an independent particle formulation of the frequency-dependent (ω) Drude–Lorentz model for solids[27,28]:

$$\hat{\varepsilon}(\omega) = 1 - \sum_{k,n} f_k^{n,n} \frac{\omega_p^2}{\omega^2 + i\eta\omega} + \sum_{k,n \neq n'} f_k^{n,n'} \frac{\omega_p^2}{\omega_{k,n,n'}^2 - \omega^2 - i\gamma\omega}, \qquad (1)$$

where $\omega_p = \sqrt{e^2 n_e / \varepsilon_0 m^*}$ is the bulk plasma frequency, $e$ is the electron charge, $n_e$ the electron density, $\varepsilon_0$ the dielectric permittivity of vacuum and $m^*$ the electron effective mass. $\hbar\omega_{k,n,n'} =$

$E_{k,n} - E_{k,n'}$ are the vertical band-to-band transition energies between occupied and empty Bloch states, labelled by the quantum numbers $(k, n)$ and $(k, n')$, respectively. $f_k^{n,n}$ and $f_k^{n,n'}$ are the oscillator strengths for the Drude and Lorentz parts and are related to the dipole matrix elements between Bloch states. $\eta$ and $\gamma$ are the Drude-like and Lorentz-like relaxation terms, which account for the finite lifetime of the electronic excitations and implicitly include the effects of the dissipative scattering.

<u>Experiments</u>: Ultrathin TiN films with thicknesses of 2, 4, 6, 8, 10 and 30 nm were grown on MgO using DC reactive magnetron sputtering. The TiN deposition was completed by sputter deposition of a titanium target in a 60% nitrogen and 40% argon environment at 5 mT with 200 W of DC power with the substrate heated to 800° C, resulting in high quality epitaxial films with low roughness[11].

The Kratos X-ray photoelectron spectrometer was used for the surface analysis of a 30 nm TiN film. A monochromatic Al K-α (1486.6 eV) X-ray source was used as the incident radiation. The wide scan for the overall spectra is performed with a scanning step of 1000 meV, while the narrow scans for the elemental peaks are performed with a scanning steps of 50 meV. The narrow scans were peak fitted using the CasaXPS software. The spectral energy scale was calibrated with the binding energy of C 1s set to 284.5 eV.

The linear optical properties of the 2, 4, 6, 8, and 10 nm films were measured using variable angle spectroscopic ellipsometry at angles of 50° and 70° for wavelengths from 400 nm to 2000 nm. A Drude-Lorentz model consisting of one Drude oscillator and one Lorentz oscillator was used to fit the measurements[11]. The charge densities in the TiN films were determined via standard Hall measurements. Using photolithography and dry etching processes, Greek crosses were fabricated from the TiN films and gold contact pads were applied. The Hall resistance was determined by measuring the voltage across two arms of the cross while a current (10 mA) passing through the other two arms generates an out-of-plane magnetic field, from which the charge density can be calculated[11].

## 3. RESULTS AND DISCUSSION

The optical properties of ultrathin TiN with thicknesses ranging from one monolayer to ten monolayers (2 nm) calculated using DFT are summarized in Figure 1. Panel b) shows the real part ($\varepsilon'$) of the dielectric function as resulting from the solution of Drude-Lorentz model of Eq. (1). Panel c) displays the thickness variation of the volume-plasmon energy $E_p=E(\omega_p)$ and of the screened-plasmon energy $E_0=E(\omega_0)$, where $\omega_0=2\pi c/\lambda_0$, and $\lambda_0$ corresponds to the wavelength where the $\varepsilon'$ changes sign. The free electron density $n_e$ (panel d) is obtained inverting the Drude-like expression for the plasma frequency $\omega_p$.

All systems exhibit a metallic character and, except for the single layer (1L), have a high carrier density of the order of $10^{22}$ cm$^{-3}$, in agreement with the experimental results (Figure 2a)[11]. Nonetheless, with decreasing thickness, the films become less metallic. The electron density $n_e$ decays almost linearly with the number of layers $N_L$ (Figure 1d) and the real part of the dielectric function (Figure 1b) decreases in magnitude as a function of $N_L$ (the imaginary part also decreases, not shown). The latter is an indication of the progressive reduction of the electronic screening in ultrathin layers. Additionally, the crossover wavelength $\lambda_0$ red shifts as the thickness decreases. This trend is in qualitative agreement with the experimental results summarized in Figure 2b, where the linear properties of epitaxial ultrathin TiN films with thicknesses of 2, 4, 6, 8, and 10 nm grown on MgO were measured using variable angle spectroscopic ellipsometry[11]. At the crossover wavelength $\lambda_0$, the imaginary part of the dielectric function $\varepsilon''$ also has a minimum. The condition $\hat{\varepsilon}(\lambda_0) \approx 0$ corresponds to the possibility of exciting a plasmon-like resonance in the visible range[21,22,26]. The difference between $E_p$ and $E_0$ for each thickness value (Figure 1c) indicates that the plasmon excitations in the optical range ($E_0$) do not involve the overall charge density of the system, as for the high energy volume plasmon ($E_p$), but only a reduced "screened" fraction ($n_e \approx 10^{21}$ cm$^{-3}$). This is the result of a complex interplay between interband and intraband transitions that effectively screen the amount of free charge that can be collectively excited in the optical range[26]. $E_p$ and $E_0$ have two evident functional trends as a function of the thickness: $E_p$ varies in the range 3.1-8.9 eV (1L-10L) with an almost ideal square root behavior (i.e. $n_e$ varies linearly with thickness), while $E_0$ is limited to a smaller variation range 1.2-2.5 eV for 1-10 layers, respectively. We further note that the corresponding values for the 3D TiN bulk system are $E_p^{bulk} = 25.4$ eV and $E_0^{bulk} = 2.7$ eV[26,29]. We conclude that (i) the increase of electron density with thickness does not linearly affect the low-energy plasmon $E_0$, (ii) interband-transition screening is active even in ultrathin films (see below); (iii) despite their metallic character these films have not reached the bulk (or thick-film) density limit, remaining well within the ultrathin regime.

The red shift of the crossover wavelength along with the reduction of the real and imaginary parts of the dielectric function when the thickness is reduced perfectly fits with a recent electromagnetic model explicitly developed to describe the confinement effects on the optical properties of ultrathin plasmonic films[17]. Here, taking advantages of first principles results, we can discuss the effect of quantum confinement in terms of the electronic structure of the ultrathin films. The electron density of states (DOS) of each TiN film close to the Fermi energy ($E_F$) is characterized by two contiguous groups of bands, labeled **1** and **2** in Figure 3, which shows the DOS spectra for the limiting cases 1L (red) and 10L (black). Bands forming the multiplet **1** have a predominant N(*2p*) character with a minor contribution from Ti(*e_g*) orbitals, while bands of group **2** mostly derive from Ti(*t_{2g}*) orbitals slightly hybridized with N(*2p*) states. The plasmonic behavior of TiN originates from intraband transitions of group-**2** bands that cross the Fermi level, giving rise to a Drude-like tail in the real part of the dielectric function and to the high-energy volume plasmon. The excitation of interband **1→ 2** transitions at lower energies gives a positive contribution to the (negative) Drude component of ε' that becomes globally positive at the crossover frequency $\lambda_0$. Thus, in a first approximation, the energy position of band **1** with respect to the Fermi level (ΔE in Figure 3) defines the energy of the screened plasmon ($E_0$). The effect of confinement actually changes the energy width and the energy position of bands **1** and **2** with respect to $E_F$. Despite the two sets of bands are clearly present in both systems, their characteristics are quite different: with decreasing thickness the surface-to-bulk ratio increases along with the number of under-coordinated surface atoms. The presence of frustrated bonds imparts a shrinking of the bandwidth of bands **1** and **2** (i.e. higher spatial electron localization), and an upshift of occupied bands (i.e. lower binding energies), which results in a decrease of ΔE (i.e. $E_0$ red-shifts). Thus, despite the same formal electron density, the reduction of thickness increases the surface contribution and the electron localization, giving an effective reduction of the free electron charges.

This qualitatively explains the modifications of the optical properties observed in the experiments varying the film thickness (Figure 2)[11]. In order to have a more quantitative comparison, we superimpose the experimental ε' spectrum (dashed black line in panel 1b) for the 2 nm film. Albeit in qualitative agreement, the crossover wavelength of 2nm TiN film is reached at a larger thickness in the experimental studies compared to the theoretical results. In particular, the theoretical value of for the 10L system, $\lambda_0$=489 nm (black thick line in Fig. 1b), is blue shifted by 96 nm with respect to the experimental value, $\lambda_0$=585 nm (black dashed line in Fig. 1b).

Two different factors resulting in this difference between experiment and theory are here considered: (i) surface oxidation and (ii) interface strain. Typically, a layer of $TiN_xO_y$ of approximately 1 – 2 nm is formed on the film surface[30], which would reduce the effective thickness of the films. X-ray photoelectron spectroscopy was conducted on a 30 nm TiN film grown on MgO to study the oxidation mechanisms on the film surface (Figure 4). The $N_{1s}$, $O_{1s}$, and $Ti_{2p}$ peaks are decomposed as shown in Figures 4b, 4c, and 4d, respectively. Three peaks can be used to fit the $N_{1s}$ spectrum, where the dominant peak at 396.8 eV is attributed to Ti-N bonds. The peak at 397.7 eV is due to the formation of an oxynitride compound, resulting in Ti-O-N bonds[31,32]. An additional weak peak in the fitting is observed at a lower energy 395.8, which may be the result of other impurities or nitrogen chemically bonded on the surface[32,33]. While evidence of $TiN_xO_y$ is also present in the $O_{1s}$ spectrum (531.2 eV), the main peak at 529.9 eV is indicative of the existence of Ti-O bonds in the films[34]. These results are also corroborated by the decomposition of the $Ti_{2p}$ peak using three pairs of doublets (2p 1/2 and 2p 3/2) for Ti-N, Ti-N-O, and Ti-O[34], implying that the films are systems consisting of TiN, $TiN_xO_y$, and $TiO_2$.

To determine how the different oxidation phases possibly modify the optical properties, nine different configurations of an oxide capping layer on a 2 nm thick film are considered. These include oxygen atoms in N substitutional sites ($O_N$), single oxygen ($O_s$) and $O_2$ molecules adsorbed on the surface (Figure 5a), and their combination. The corresponding ε' plots are collected in panels b) and c). The dielectric function of ideal 2nm thick layer (i.e. 10L, black thick line) and the corresponding experimental results (black dashed line) are included for comparison.

Since the metallic bands at Fermi is mainly due to Ti-derived states, the coordination of Ti atoms is crucial in the resulting optoelectronic properties of the system. If a single or a few O atoms substitute surface nitrogens, this does not substantially change the Ti coordination leaving the electronic structure almost unchanged (Figure 5b). Rather, $O_N$ is an aliovalent impurity that acts as an n-type dopant, which increases the electron density. Consequently, a blue shift in the crossover wavelength is observed, contrary to experimental results. Conversely, single oxygen atoms on the surface have little effect on the optical properties (Figure 5b). On the contrary, $O_2$ molecules do cause a red shift of the crossover wavelength, tending towards the experimental value (Figure 5c). The $O_2$ molecules change the coordination of the Ti atoms on the surface to resemble the configuration for $TiO_2$. As the amount of oxygen is increased, surface Ti easily coordinates with oxygen forming Ti-O bonds at the surface, reducing the overall thickness of pure TiN and leading to the slower blue shift of the crossover wavelengths observed experimentally. To prove this statement we consider, as key example, the case where $3O_N$ and $2O_2$ are included in the system (labeled $3O_N+2O_2$, Figure 6). After atomic relaxation, the surface layer displays a remarkable structural rearrangement, where the Ti atoms may have three different kinds of coordination: (i) a partially detached oxide layer forms on surface, where Ti is bonded to O atoms only. This is in agreement with the Ti-O feature revealed in the XPS data (Figure 4c). (ii) In the central region of the slab Ti is 6-fold coordinated with N, and forms the TiN core of the film (see also main peak in Figure 4b). (iii) In the intermediate region N and O coexist and Ti atoms are coordinated with both species giving rise to mixed Ti-N-O phases. This is coherent with the appearance of Ti-N-O features in $N_{1s}$ and $O_{1s}$ spectra.

The effects on the electronic structure are evident in Figure 6b, where the comparison with the DOS of ideal 10L system is presented. The total DOS of $3O_N+2O_2$ system is reminiscent of the original bands **1 - 2** and the system maintains a metallic character, but the presence of oxygen (shaded area) includes new localized states across the Fermi level. This has a twofold effect: first, it reduces the effective thickness of TiN; second, it introduces further O-derived states that contribute to interband transitions, further shifting the crossover wavelength toward the red. The oxide layer, however, is not sufficiently thick to exhibit the characteristic dielectric features of $TiO_2$. Although the exact microscopic reconstruction of the surface strictly depends on the growth process, this example is sufficient to give optical properties in quantitative agreement with experimental findings (Fig. 5b).

Since experimental TiN samples are not free standing but grown on MgO substrate, the effect of a substrate induced strain is included in the simulations and compared to experiments. The lattice mismatch would introduce a strain in the film, changing the in-plane lattice parameter of TiN. In order to investigate this effect without including other structural modifications derived by the explicit description of the substrate, we simulated a set of free standing 10-layer-thick TiN films with varying in-plane lattice constants, which span the range [-3% ÷ +6%], with respect to the equilibrium lattice parameter ($a_0^{TiN}$=4.24 Å), as shown in Figure 7. As the lattice parameter is reduced (compressive strain), the crossover wavelength blue shifts; the opposite happens when the lattice parameter is increased (tensile strain). To reach the experimental value for the crossover wavelength, a tensile strain of ~6% is required. Although ultrathin films may sustain large amounts of strain due to interface mismatch, the latter condition (+6%) does not fit the experimental situation as the actual mismatch between TiN and MgO would result in a compressive strain of only ~0.7% ($a_0^{MgO}$=4.21 Å). Thus, even though in principle strain may affect the optical properties of ultrathin films, in the present case oxidation seems to be the most important cause of the differences between the ideal TiN model and measured sample with the same formal thickness.

To support these results, the substrate induced strain is experimentally determined by high-angle annular dark field imaging scanning transmission electron microscopy (HAADF STEM) of a 2 nm TiN film grown on MgO, as shown in Figure 8. The observed contrast at the TiN/MgO interface signifies the formation of misfit dislocations, indicating that the film has relaxed and has no substrate induced strain which may have influenced the optical properties. The strain relaxation mechanism is typically dependent on the growth process of the material. If the first few monolayers of the TiN grown on MgO form epitaxial islands having the same crystal orientation, the coalescence of the islands would create dislocations contributing to strain relaxation. Once the islands merge, a smooth, continuous, epitaxial film continues to grow[35,36]. This is consistent with the theoretical findings that the oxidation plays the main role in the difference between the simulations for pure TiN and experimental work. However, as the thickness decreases, the films may remain strained and exhibit a stronger effect on the changes in the optical properties.

## 4. CONCLUSION

We have conducted a detailed theoretical and experimental study on the plasmonic behavior of ultrathin (down to a few atomic layers) TiN films to determine the role of thickness, surface oxidation and interface strain on TiN optical properties. As the thickness is reduced, the metallic character is also decreased as a result of quantum confinement effects, which increase the contribution from under-coordinated surface atoms leading to an increased electron localization. The atomistic analysis of the experimentally grown TiN thin films indicates a non-negligible deviation from the ideal TiN structure on the surface. We show that the fabricated films' composition can be described as a combination of three different phases: titanium nitride, titanium oxynitride, and titanium dioxide. The surface oxidation of the films, particularly affecting the Ti coordination, has a significant influence on the optical properties of the TiN films, which red shifts the crossover wavelength with respect to the ideal case with the same formal thickness. Additionally, although there is no strain observed in the current films studied, thinner films may exhibit a change in the lattice parameter, which would further affect the optical properties.

The investigated ultrathin films remain highly metallic and are characterized by plasmonic excitations in the visible wavelength range. More importantly, the fabricated ultrathin films' optical response can be adjusted by controlling their structural and compositional parameters, such as thickness, strain, and oxidation during the growth. The observed plasmonic properties, in combination with the confinement effects, make TiN ultrathin films a promising material for the realization of plasmonic metasurfaces with enhanced nonlinearities and electrical tunability. Thus, ultrathin plasmonic TiN films can be profitably exploited in flexible optoelectronics, e.g. smart wearable, or biomedical implantable devices.

## 5. ACKNOWLEDGMENTS

This work was supported from the NSF OP grant DMR–1506775. The authors would like to acknowledge Dr. Dmitry Zemlyanov for help with XPS characterization and Dr. Sergei Rouvimov for help with TEM characterization.

**Figures**

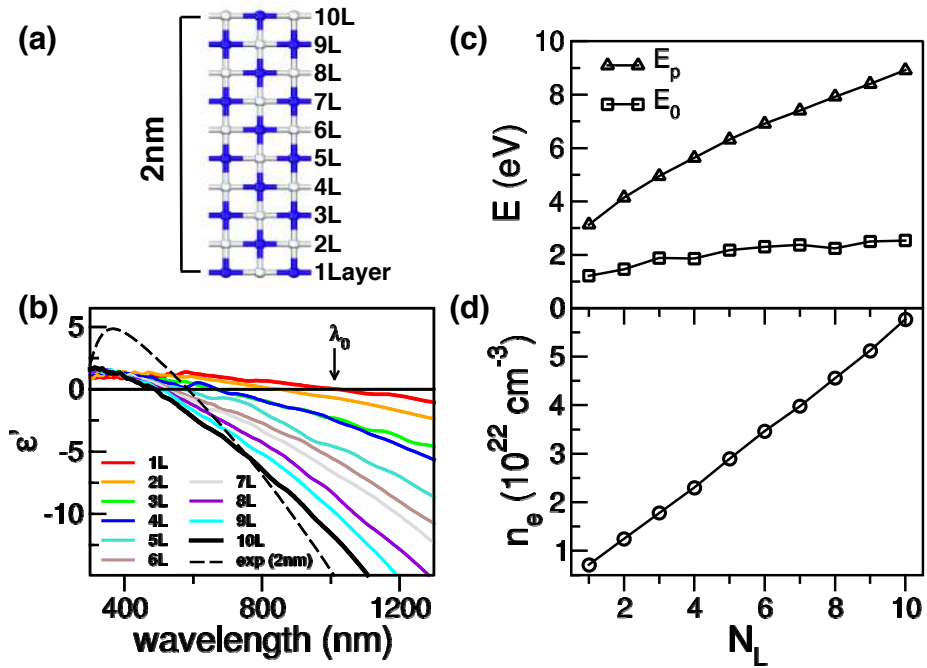

**Figure 1** (a) Side view of 1 to 10 layer thick TiN(100) slabs used in the simulations. (b) Real part of the dielectric function as a function of the number of layers $N_L$. Experimental spectrum (black dashed line) is included for comparison. (c) Bulk plasmon ($E_p$), screened plasmon ($E_0$) energies and (d) free electron charge density ($n_e$) for 1-10 layer films.

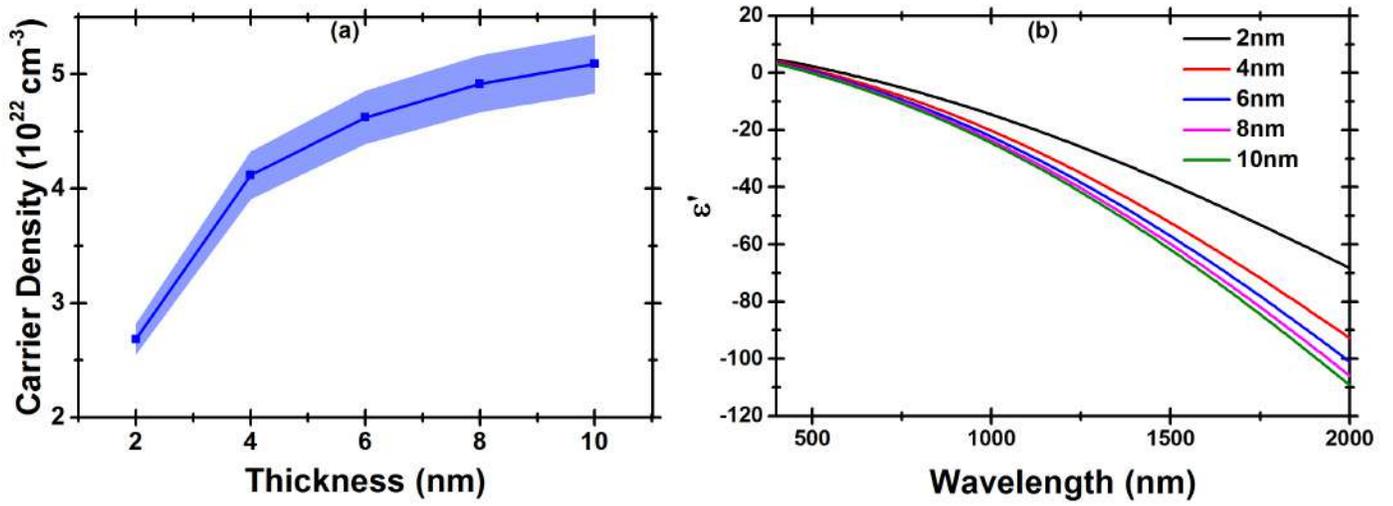

**Figure 2** (a) Charge density obtained from Hall measurements and (b) real part of the dielectric function from ellipsometry measurements for increasing TiN film thicknesses.

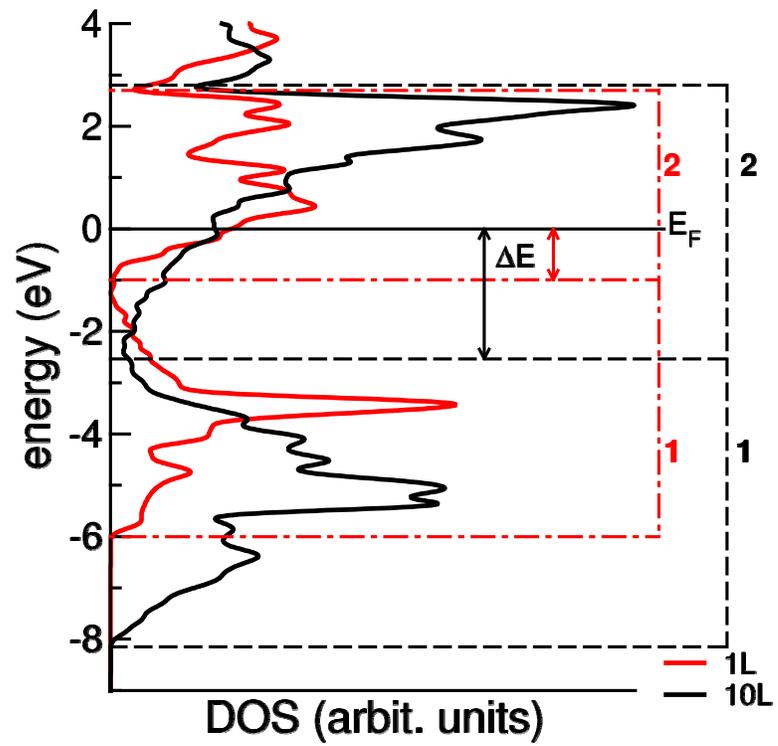

**Figure 3**. Simulated DOS plots for 1L (red) and 10L (black) systems. Dashed lines remark energy position with respect to the Fermi level ($E_F$) and the energy width of the two main groups of bands (**1**, **2**) responsible for crossover wavelength in the visible range ($\lambda_0$).

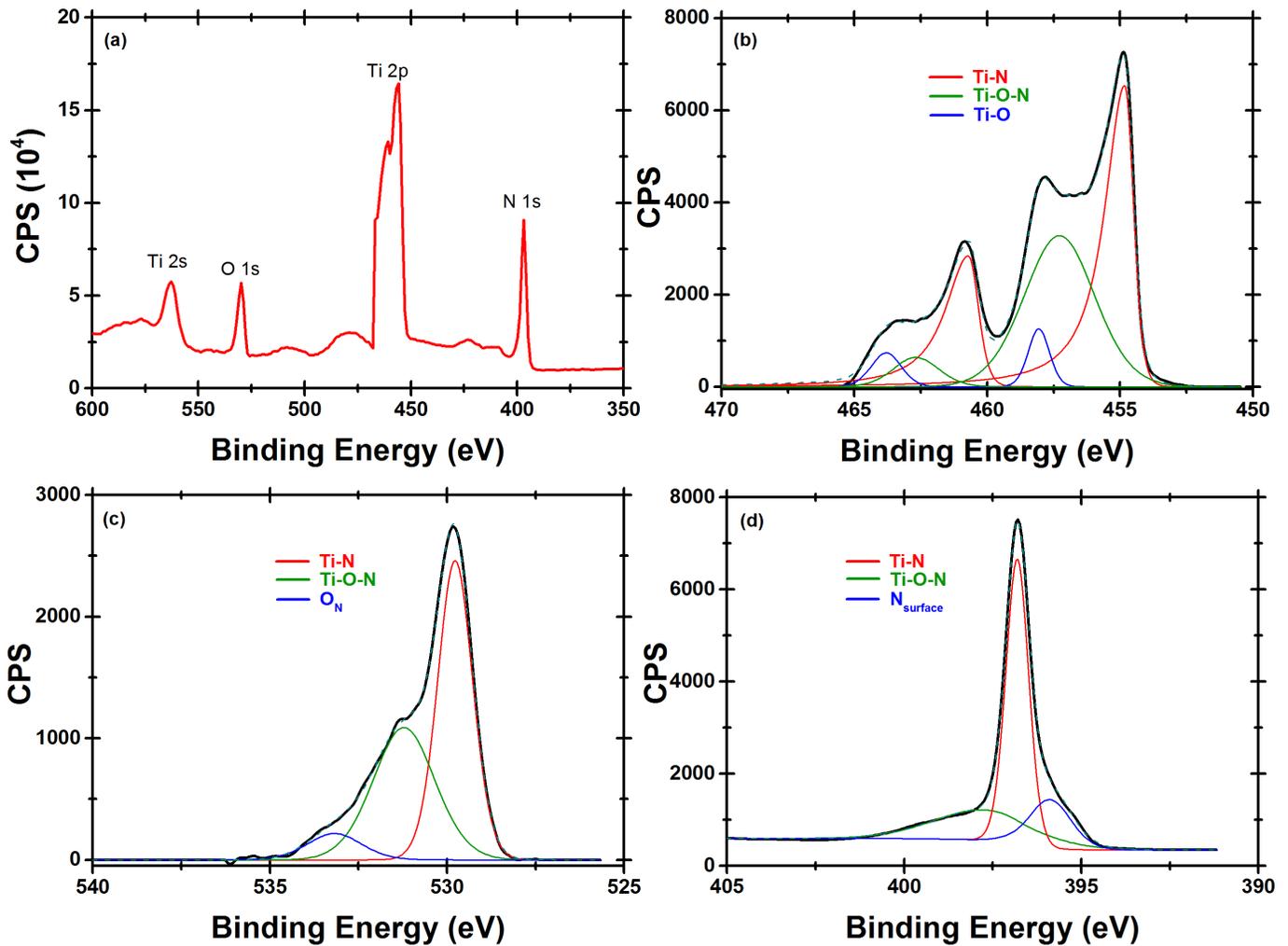

**Figure 4.** (a) XPS spectrum of a partially oxidized 30nm TiN film grown on MgO by DC magnetron sputtering. The (b) N1s, (c) O1s, and (d) Ti2p peaks are fitted to determine the local chemical environment. The black curve represents the actual XPS data collected, while the blue dashed curve is the generated envelope curve from peak fitting.

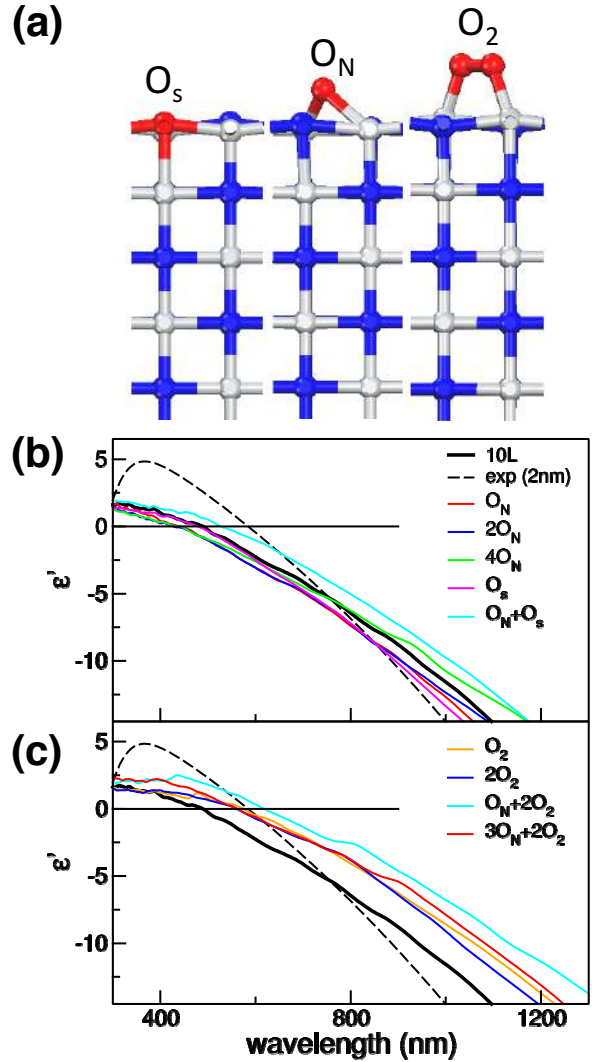

**Figure 5.** (a) Configurations for single oxygen atom in a nitrogen substitutional site ($O_N$), single oxygen atom on the surface ($O_s$) or diatomic oxygen molecule on the film surface ($O_2$). Real part of the dielectric function of oxidized TiN film including (b) single atoms $O_N$ and $O_s$; and (c) $O_2$ molecule. Different combinations of these three main configurations are modeled to determine the effect of oxygen in the films. Theoretical ideal model (10L) and the corresponding experimental spectra are reported in panels b and c for direct comparison.

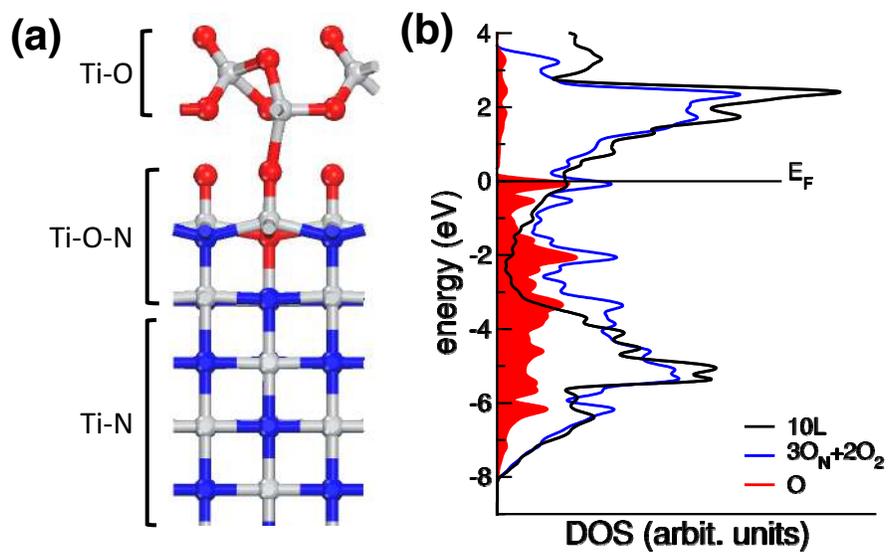

**Figure 6.** (a) Relaxed geometry of $3O_N+2O_2$ configuration. (b) Total (blue line) and O-projected (shaded area) density of states of $3O_N+2O_2$ configuration compared to the ideal unoxidized 10L case (black line). Zero energy reference is set to the Fermi level of the two systems.

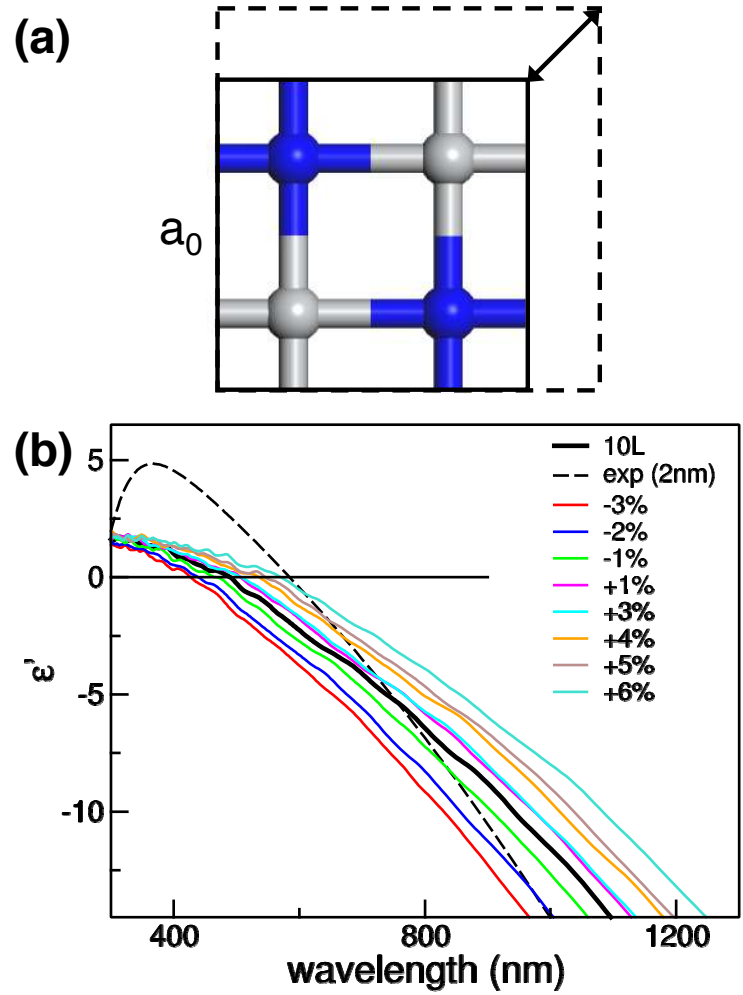

**Figure 7.** (a) In-plane strain model. (b) Simulated real part of the dielectric function for 10L TiN films, with varying amount of applied lateral strain. Theoretical ideal unstrained model (10L) and corresponding experimental spectra are reported in panels b for direct comparison.

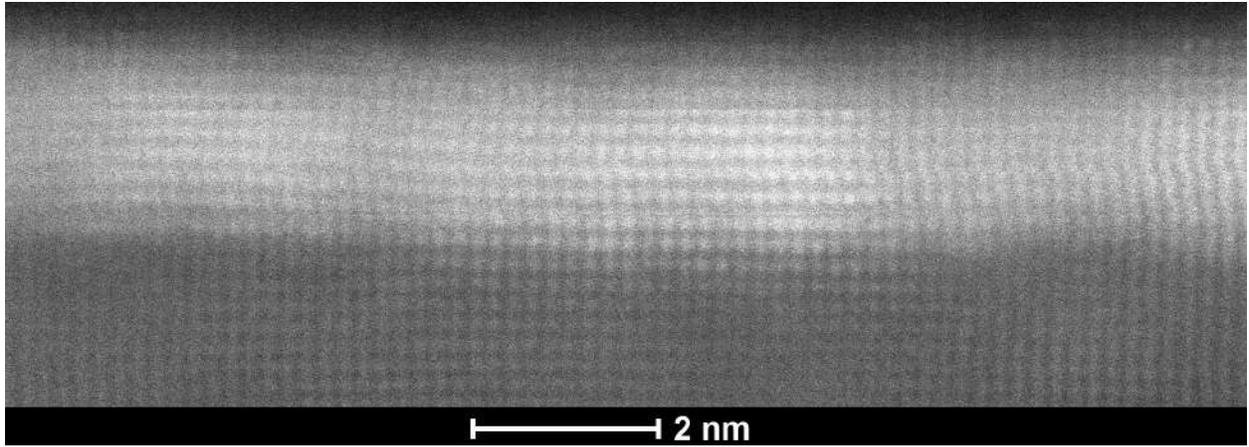

**Figure 8.** HAADF STEM image of a 2 nm TiN film on an MgO substrate. The contrast at the film/substrate interface is due to misfit dislocations, causing the film to relax.

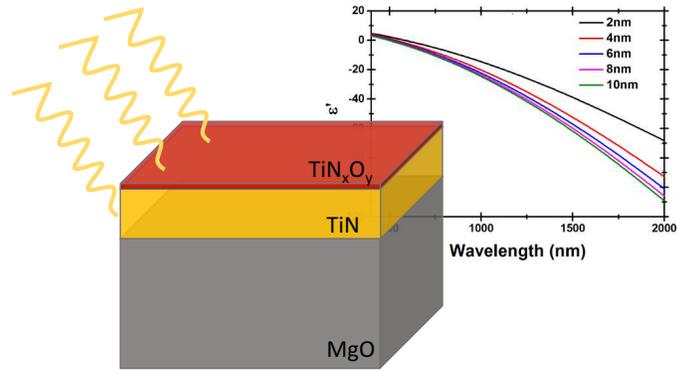

TOC graphic